\documentclass[aps,prd,reprint,superscriptaddress]{revtex4-2}
\usepackage{amsmath}
\usepackage{siunitx}
\DeclareSIUnit{\gauss}{G}
\usepackage{xcolor}
\usepackage{graphicx}
\usepackage[normalem]{ulem}  
\newif\ifshowcomments
\showcommentstrue            
\NewDocumentCommand{\cmtcite}{o m}{\mbox{\IfValueTF{#1}{\cmtoldcite[#1]{#2}}{\cmtoldcite{#2}}}}%
\NewDocumentCommand{\newcommenter}{m m m}{%
  \definecolor{cmt#2}{HTML}{#3}%
  \NewDocumentCommand{#1}{+o +m}{%
    \ifshowcomments
      {\color{cmt#2}%
       \IfValueT{##1}{{\let\cmtoldcite\cite\let\cite\cmtcite\sout{##1}}\ }%
       {\setlength{\fboxsep}{1pt}\colorbox{cmt#2}{\color{white}\sffamily\bfseries\scriptsize #2}}%
       \ ##2}%
    \else
      \IfValueT{##1}{##1}%
    \fi
  }%
}
\newcommenter{\Morgan} {M}  {C000C0} 
\newcommenter{\Ola}    {O}  {7B1FA2} 
\newcommenter{\Diego}  {D}  {1565C0} 
\newcommenter{\Marek}  {K}  {6D4C41} 
\newcommenter{\Sara}   {S}  {B8860B} 
\newcommenter{\Bertran}{B}  {E65100} 
\newcommenter{\Zoi}    {Z}  {827717} 
\newcommenter{\Wen}    {W}  {2E7D32} 

\usepackage{color}
\definecolor{mygreen}{rgb}{0,0.5,0}
\definecolor{babyblue}{RGB}{220,238,250}
\definecolor{mygrey}{rgb}{0.5,0.5,0.5}
\definecolor{myred}{rgb}{0.75,0,0}
\definecolor{myblue}{rgb}{0,0,0.75}
\definecolor{mymagenta}{cmyk}{0,1,0,0.12}
\definecolor{mycyan}{cmyk}{1,0,0,0.12}
\definecolor{myorange}{rgb}{1.,0.5,0}
\definecolor{myviolet}{rgb}{0.6,0.15,0.6}
\definecolor{mybrown}{cmyk}{0,0.50,1,0.41}

\makeatletter
\newcommand{\figurepanellabel}[2]{%
  \begingroup
  \protected@edef\@currentlabel{\thefigure(#1)}%
  \label{#2}%
  \endgroup
}
\makeatother

\begin{document}


\title{Projected sensitivity of a spinor Bose-Einstein condensate comagnetometer \\ to proton monopole-dipole interactions}




\newcommand{\ICFO}{ICFO -- Institut de Ciències Fotòniques, The Barcelona Institute of Science and Technology, 08860 Castelldefels (Barcelona), Spain}
\newcommand{\IFAE}{Institut de F\'{i}sica d'Altes Energies (IFAE), The Barcelona Institute of Science and Technology, Campus UAB, 08193 Bellaterra (Barcelona), Spain}
\newcommand{\ICREA}{Instituci\'o Catalana de Recerca i Estudis Avan\c{c}ats (ICREA),
Passeig Llu\'is Companys 23, 08010 Barcelona, Spain}

\author{Wenjing Zhou}
\affiliation{\ICFO}
\author{Diego Blas}
\affiliation{\IFAE}
\affiliation{\ICREA}
\author{Morgan W. Mitchell}
\affiliation{\ICFO}
\affiliation{\ICREA}



\newcommand{\CP}{\mathit{CP}}
\newcommand{\CPT}{\mathit{CPT}}
\renewcommand{\um}{\si{\mu m}}
\newcommand{\SO}{S_\Omega}
\newcommand{\srcint}{\mathcal{G}}  


\date{\today}

\begin{abstract}
Light spin-zero bosons with scalar and pseudoscalar couplings mediate a parity- and time-reversal-violating monopole-dipole interaction between unpolarized matter and polarized spins. For proton spins, the interaction has been probed directly with atoms only at ranges $\lambda \gtrsim 5\, \si{mm}$. At shorter ranges, laboratory limits come mostly from neutron-dominated nuclei, through nuclear-spin models. We analyze the prospects for a direct search with a single-domain $^{87}$Rb spinor Bose-Einstein condensate, whose $F=1$ and $F=2$ manifolds act as two co-located magnetometers. They precess in opposite directions in a magnetic field but in the same direction under the exotic interaction. The sum of their precession angles thus suppresses field noise by a measured $44\, \si{dB}$, and a weighted sum cancels the field altogether. Starting from the magnetometry [Proc. Natl. Acad. Sci. U.S.A. 119, e2115339119 (2022)] and comagnetometry [Phys. Rev. Lett. 124, 170401 (2020)] already demonstrated with the existing apparatus, we project its sensitivity as a function of the source geometry and of in-progress upgrades to the atom loading, the trap beams and the magnetic environment. With an insulating bismuth germanate source whose spherical top comes within $50\, \um$ of the center of the condensate, one week of data would reach $|g_s^N g_p^p|\approx 2\times10^{-22}$ with these improvements and $3\times10^{-24}$ at the projection-noise limit ($\lambda=100\, \um$, 95\% C.L.), more than three and five orders of magnitude below the existing laboratory limits. The improvement covers boson masses of 0.3--$50\, \si{meV}/c^2$ and 0.09--$70\, \si{meV}/c^2$, respectively.
\end{abstract}

\maketitle

\begin{figure*}[t]
\centering
\includegraphics[height=35mm]{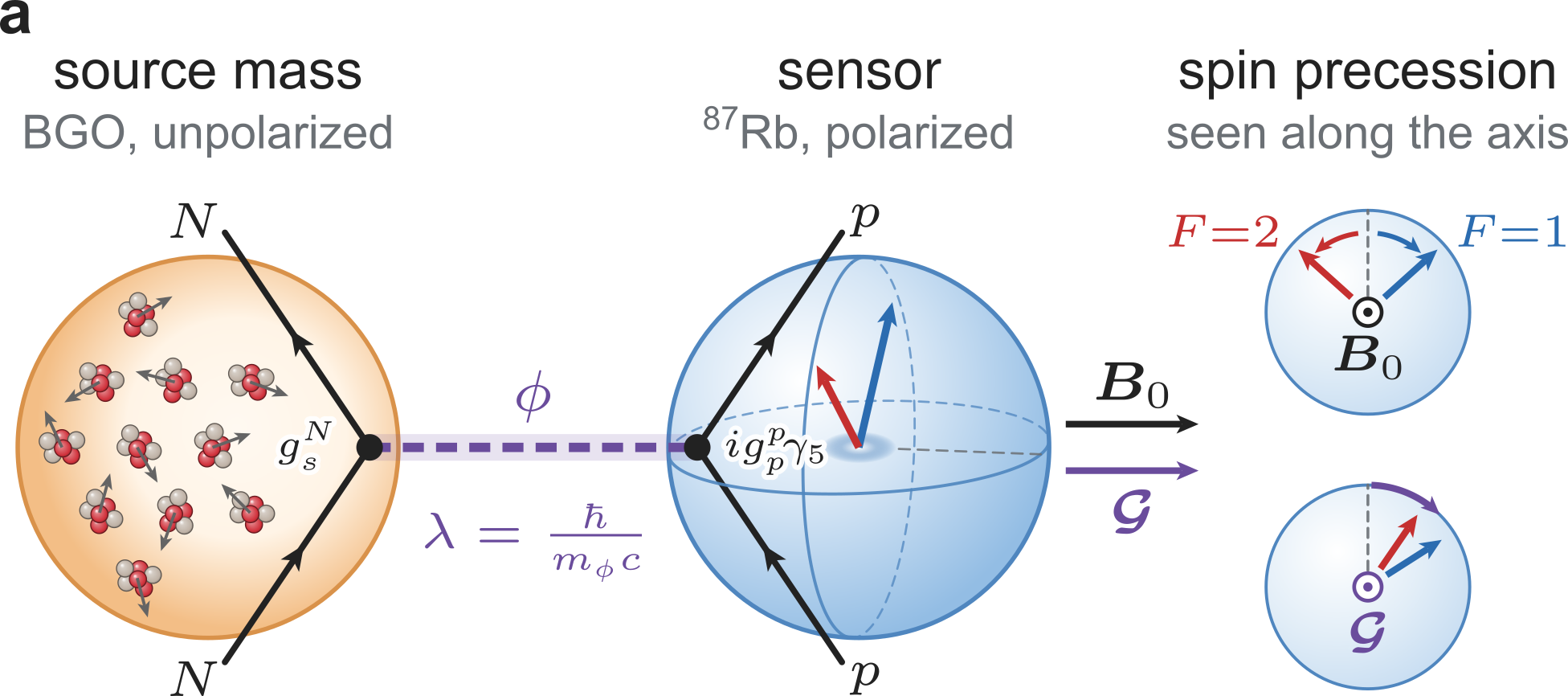}\hspace{6mm}
\includegraphics[height=35mm]{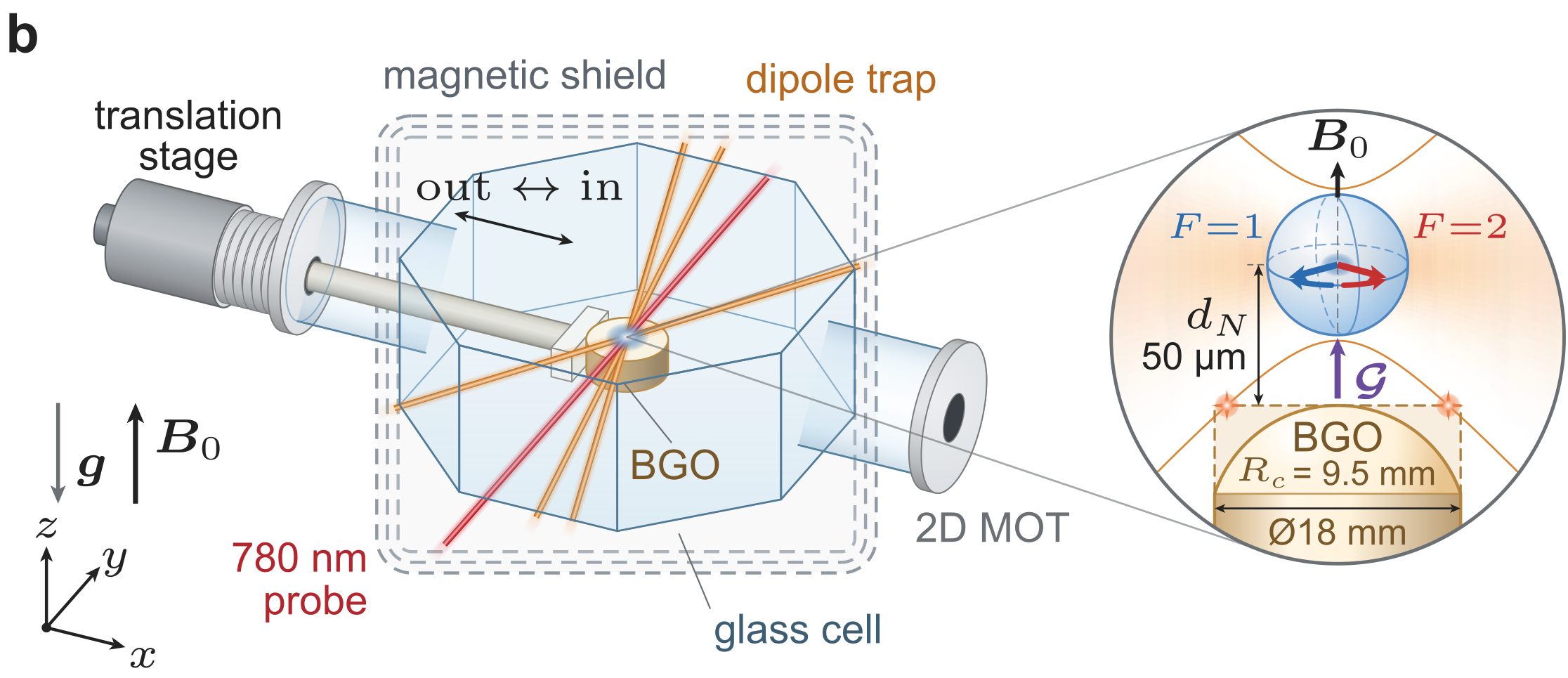}\\[1.5mm]
\includegraphics[width=\textwidth]{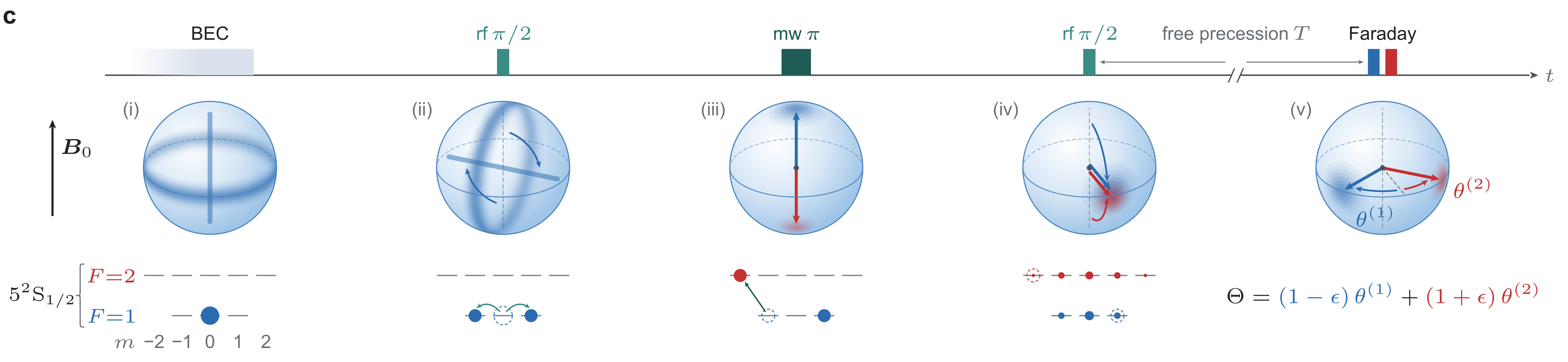}
\caption{(a)~Monopole-dipole interaction between a nucleon $N$ of the source and the valence proton $p$ of $^{87}$Rb, mediated by a boson $\phi$ of range $\lambda$. Seen along $\boldsymbol{\srcint}$ and $\boldsymbol{B}_0$ (right), the bias field turns $F=1$ and $F=2$ in opposite directions, and the source turns both the same way.
(b)~Apparatus. A translation stage moves the BGO source in and out under the condensate, apex $d_N=50\, \um$ below the atoms. Inset: a focused beam widens by diffraction. The spherical top falls away below it, whereas a flat top (dashed) would cut it.
(c)~One experimental shot. The condensate starts in $|1,0\rangle$, with no net spin; the bar marks its alignment axis (i). An rf $\pi/2$ pulse tips this axis into the transverse plane (ii), and a microwave $\pi$ pulse from $|1,-1\rangle$ to $|2,-2\rangle$ leaves $F=1$ along $\boldsymbol{B}_0$ and $F=2$ against it (iii). A second rf $\pi/2$ pulse tips both spins into the transverse plane, where they precess (iv). Two Faraday pulses then read out their azimuths $\theta^{(1)}$ and $\theta^{(2)}$, which give the weighted sum $\Theta$ of Sec.~\ref{Subsec: Observable} (v).}
\figurepanellabel{a}{Fig: 1a}
\figurepanellabel{b}{Fig: 1b}
\figurepanellabel{c}{Fig: 1c}
\end{figure*}

\section{\label{Sec: Intro} Introduction}
The Higgs boson is the only elementary spin-zero particle observed to date \cite{Aad_2012, Chatrchyan_2012, Aad_2013}, yet many extensions of the Standard Model predict new spin-zero bosons that are orders of magnitude lighter. Among the best-motivated examples are the QCD axion \cite{Peccei_1977, Weinberg_1978, Wilczek_1978}, the axion-like particles of string compactifications \cite{Svrcek_2006, Arvanitaki_2010}, the relaxion \cite{Graham_2015}, and light dilatons and moduli \cite{Damour:2010rp, Damour:1994zq, Dimopoulos:1996kp}, several of them compelling dark-matter candidates \cite{Preskill_1928, Abbott_1982, Dine_1982, Sikivie_2021}. The exchange of such a boson $\phi$, of mass $m_{\phi}$, produces a force of range $\lambda= \hbar /m_\phi c$. If $\phi$ has both scalar and pseudoscalar couplings to fermions $\psi$, $\mathcal{L}_\phi=\phi \sum_\psi \bar{\psi}\left(g_s^\psi+i \gamma_5 g_p^\psi\right) \psi$ ($\hbar= c= 1$), it mediates a monopole-dipole interaction \cite{Moody_Wilczek_1984, Dobrescu_2006, Fadeev_2019} between an unpolarized particle $X$, coupling through $g_s^X$, and a fermion $Y$ of mass $m_Y$ and Pauli vector $\boldsymbol{\sigma}_Y$, coupling through $g_p^{Y}$. Following the convention of Ref.~\cite{Cong_2025}, with dimensionless couplings, the monopole-dipole potential is
\begin{equation}
V(\boldsymbol{r})=-\frac{g_s^X g_p^Y \hbar^2}{8 \pi m_Y} \boldsymbol{\sigma}_Y \cdot \hat{\boldsymbol{r}}\left(\frac{1}{\lambda r}+\frac{1}{r^2}\right) e^{-r / \lambda},
\label{Eq: MW}
\end{equation}
where $\hat{\boldsymbol{r}}$ points from the unpolarized source particle $X$ to the polarized test spin $Y$. The unpolarized particle thus acts on the spin like a magnetic field along $\hat{\boldsymbol{r}}$, produced through $g_s^X$ and felt through $g_p^Y$ instead of the magnetic moment, so that atomic magnetometers and comagnetometers are well suited to detect it.

Because $\boldsymbol{\sigma}_Y\cdot\hat{\boldsymbol{r}}$ is odd under parity and time reversal, a nonzero product $g_s^X g_p^Y$ makes the interaction violate both symmetries and, by the $\CPT$ theorem, $\CP$ as well. Its detection would be significant in its own right: all $\CP$ violation observed so far is consistent with the quark-mixing phase of the Standard Model, which is far too weak to explain the matter--antimatter asymmetry of the Universe \cite{Gavela_1994, Huet_1995, Dine_2003}. For the QCD axion, this violation is set by the effective QCD vacuum angle $\bar{\theta}$, which the electric dipole moment of the neutron bounds below about $10^{-10}$ \cite{Abel_2020}. The product is then far below the reach of present laboratory experiments \cite{Bertolini_2021, DiLuzio_2023, OHare_2020}, which are therefore sensitive mainly to other light bosons, such as axion-like particles, for which $g_s$ and $g_p$ are free parameters. Laboratory searches constrain the couplings of the electron \cite{Wineland_1991, Youdin_1996, Ni_1999, Hammond_2007, Heckel_2008, Hoedl_2011, Terrano_2015, Crescini_2017, Rong_2018, Lee_2018, Stadnik_2018, Crescini_2022, Liang_2023} and of the neutron \cite{Wineland_1991, Venema_1992, Youdin_1996, Baebler_2007, Serebrov_2009, Voronin_2009, Petukhov_2010, Fu_2011, Bulatowicz_2013, Chu_2013, Tullney_2013, Guigue_2015, Lee_2018, Feng_2022, Ayres_2023} over many decades of $\lambda$. For proton spins, the interaction has been probed directly only at $\lambda \gtrsim 5\, \si{mm}$, with Rb and Cs atoms, whose nuclear spin is carried by a valence proton \cite{Kimball_2015, Kimball_2017}; below, proton limits are inferred from $^{3}$He and $^{129}$Xe data \cite{Stadnik_2015, Cong_2025} through the proton spin content of these nuclei, which is small for $^{3}$He and, for $^{129}$Xe, differs between nuclear models by more than an order of magnitude \cite{Kimball_2015, Stadnik_2015}.

A direct search below a few millimeters needs polarized protons within about $\lambda$ of a dense source, since the force decays as $e^{-r/\lambda}$. Many searches cited above use atomic vapors or polarized noble gases in cells, where the wall sets the minimum source-spin distance and the atoms sample the whole cell during a measurement. A single-domain spinor Bose-Einstein condensate (SBEC) avoids both limitations. It is a motionless ensemble of about $10^5$ atoms, with a Thomas-Fermi radius of $\approx 7\, \um$, in which all atoms share one spatial mode~\cite{Palacios_2022}. Condensates have been held within micrometers of a surface \cite{Rychtarik2004} and brought even closer \cite{Bender2009, Bender2014}. The two hyperfine manifolds of the condensate naturally form a comagnetometer \cite{Gomez_2020}: a magnetic field turns $F=1$ and $F=2$ in opposite directions, whereas a coupling to the nuclear spin, which in $^{87}$Rb is carried by the valence proton, turns both the same way [Fig.~\ref{Fig: 1a}]. In Ref.~\cite{Gomez_2020}, the sum of the two precession angles suppressed magnetic noise by $44\, \si{dB}$. A coupling to the nuclear spin would instead add up in this sum, and a slightly different weighting of the two angles cancels the field altogether (Sec.~\ref{Sec: Theory}). Used as a magnetometer, a single-domain SBEC was the first sensor to reach an energy resolution per bandwidth below $\hbar$ \cite{Mitchell_2020, Palacios_2022}.

Here we analyze a direct search for the proton monopole-dipole interaction with an SBEC comagnetometer and a source mass of bismuth germanate (BGO), a dense insulator, placed under the condensate. With a margin for alignment tolerances, the apex of its spherical top could come within $150\, \um$ of the atoms with the present horizontal trap and probe beams, and within $50\, \um$ with refocused ones. Moving the source in and out in alternate shots cancels the systematic effects that do not depend on its position [Fig.~\ref{Fig: 1b}]. We project the sensitivity in two scenarios built on the demonstrated performance of the apparatus: U, after three upgrades in preparation (a magnetic shield, faster loading from a two-dimensional magneto-optical trap, and refocused trap and probe beams), and Q, at the quantum projection-noise limit. Even the apparatus as demonstrated in Ref.~\cite{Gomez_2020} would improve on the existing proton limits between $20\, \um$ and $300\, \um$, by a factor of 50 at $\lambda=100\, \um$. Scenario U would improve on them for $\lambda$ between $4\, \um$ and $0.6\, \si{mm}$, by $5\times10^3$ at $100\, \um$, and scenario Q between $3\, \um$ and $2\, \si{mm}$, by $3\times10^5$. Although the apex sits $50\, \um$ from the atoms, the improvement extends to ranges about ten times shorter (Appendix~\ref{App: Scans}).

The search rests on four elements, each discussed below. The weighted sum of the two phases cancels any magnetic field common to both manifolds, with weights set by the $g$ factors of $^{87}$Rb and known to one part in $10^6$ (Sec.~\ref{Subsec: Observable}). The BGO source carries no thermal currents, and hence none of the Johnson noise that a metal would produce at $50\, \um$ from the atoms (Sec.~\ref{Sec: Sources}). Its spherical top falls away from the trap and probe beams as they widen, so that it can come much closer than a flat face. Finally, reversing $\boldsymbol{B}_0$ between blocks of in-out pairs also cancels, by symmetry, the systematic effects that reverse with the field (Sec.~\ref{Subsec: Sequence}).

Section~\ref{Sec: Theory} describes the principle of the measurement, Sec.~\ref{Sec: Sensor} the sensor, the source mass and the measurement sequence, Sec.~\ref{Sec: Sensitivity Reach} the projected sensitivity, and Sec.~\ref{Sec: Conclusion} the outlook. Appendix~\ref{App: Scans} shows how the reach depends on the design.

\section{Principle of the measurement}\label{Sec: Theory}
\subsection{Signal}\label{Subsec: Signal}
Integrating Eq.~(\ref{Eq: MW}) over a source of nucleon density $n(\boldsymbol{r}')$, with equal scalar couplings to protons and neutrons, $g_s^p=g_s^n\equiv g_s^N$, and neglecting the coupling to electrons \cite{Kimball_2017, Cong_2025}, gives the interaction of a proton, $\mathcal{H}=-(g_s^Ng_p^p\hbar^2/8\pi m_p)\,\boldsymbol{\sigma}\cdot\boldsymbol{\srcint}$, with the source integral
\begin{equation}
\begin{aligned}
\boldsymbol{\srcint} & =\int d^3 r'\, n(\boldsymbol{r}')\, \hat{\boldsymbol{r}}\left(\frac{1}{\lambda r}+\frac{1}{r^2}\right) e^{-r / \lambda} \\
& =-\nabla_{p} \int d^3 r'\, \frac{n(\boldsymbol{r}')\, e^{-r / \lambda}}{r}.
\end{aligned}
\label{Eq: Pseudomagnetic Field}
\end{equation}
Here $\boldsymbol{r}=\boldsymbol{r}_{p}-\boldsymbol{r}'$ points from the source to the proton, $r=|\boldsymbol{r}|$, and $\nabla_{p}$ is the gradient with respect to $\boldsymbol{r}_p$. The source integral has units of inverse area and depends only on $\lambda$ and on the source. For the proton spin, $\boldsymbol{s}_p=\hbar\boldsymbol{\sigma}/2$, the interaction is $\mathcal{H}=\boldsymbol{\Omega}_p\cdot\boldsymbol{s}_p$, with $\boldsymbol{\Omega}_p=\Omega_p\hat{\boldsymbol{\srcint}}$ and
\begin{equation}
    \Omega_p=-\frac{g_s^N g_p^p \hbar \srcint}{4 \pi m_p}.
    \label{Eq: Omega_p}
\end{equation}
The source acts on the proton spin like a magnetic field along $\boldsymbol{\srcint}$, about which a free proton would precess at the rate $\Omega_p$. A magnetometer, which senses a field through the spin precession it drives, is therefore a natural detector.

In $^{87}$Rb, the proton spin projects onto the nuclear spin as $\boldsymbol{s}_p\rightarrow \sigma_p \boldsymbol{I}$, where $\sigma_p$ is the proton spin content of the nucleus. The nuclear spin, in turn, projects onto the spin of the hyperfine manifold $F=f$ as $\boldsymbol{I}\rightarrow c_f \boldsymbol{F}^{(f)}$ \cite{Kimball_2013, Kimball_2015}, giving
\begin{equation}
\mathcal{H}^{(f)}=c_f \sigma_p \boldsymbol{\Omega}_p \cdot \boldsymbol{F}^{(f)}, \quad c_1=\frac{5}{4}, \quad c_2=\frac{3}{4}.
\label{Eq: Projection_I}
\end{equation}

We take $\sigma_p=1/3$ from the Schmidt model, in which the nuclear spin of $^{87}$Rb is carried by a single valence $\pi p_{3/2}$ proton \cite{Schmidt_1937, Kimball_2015}. This model overestimates the nuclear magnetic moment, $3.79\, \mu_N$ against the measured $2.75\, \mu_N$ \cite{Stone_2005}. Semi-empirical models assign the difference to a spin polarization of the other nucleons and find $\sigma_p=0.18$--0.26 \cite{Engel_1989, Flambaum_2006, Stadnik_2015}. Reference~\cite{Kimball_2015} showed, however, that these models cannot reliably predict this polarization. For odd-proton nuclei, moreover, they omit a velocity-dependent spin-orbit correction to the orbital magnetic moment of the proton. For the valence protons of Rb, Ref.~\cite{Kimball_2015} concluded that the Schmidt value is reliable to within a factor of two. With the semi-empirical values, our limits would be 1.3--1.8 times weaker (Appendix~\ref{App: Scans}). The models of Ref.~\cite{Flambaum_2006} also give the neutrons a small spin content, $\sigma_n=0.07$--0.08, so that strictly speaking the measurement bounds $\sigma_p g_p^p+\sigma_n g_p^n$.

\subsection{Observable}\label{Subsec: Observable}
Magnetic-field drifts and fluctuations, however, will overwhelm this signal, even inside a magnetic shield. The source may also carry a magnetic field of its own, from its magnetization or its thermal currents. Many of the most sensitive searches therefore use comagnetometers, which compare two spin species in the same volume \cite{Cong_2025, Terrano_2022, Safronova_2018}. Both species see the same magnetic field, so that a suitable combination of their phases cancels it but keeps an exotic coupling that is not proportional to their magnetic moments. Noble-gas comagnetometers combine $^{3}$He and $^{129}$Xe \cite{Tullney_2013} or $^{129}$Xe and $^{131}$Xe \cite{Bulatowicz_2013, Feng_2022}. Comagnetometry with a single atomic species has been demonstrated in both $^{87}$Rb vapor cells \cite{Wang_2020} and Bose-Einstein condensates \cite{Gomez_2020}, using the two ground-state hyperfine manifolds, $F=1$ and $F=2$.

A magnetic field couples mainly to the electron spin $\boldsymbol{S}$, and the source, through the proton, to the nuclear spin $\boldsymbol{I}$. Within manifold $f$, $\boldsymbol{S}\rightarrow s_f \boldsymbol{F}^{(f)}$, with $s_{1,2}= \mp 1/4$ \cite{Kimball_2013, Kimball_2015}. The electron spin is therefore antiparallel to the total spin in $F=1$ and parallel in $F=2$, whereas the nuclear spin is parallel in both [Eq.~(\ref{Eq: Projection_I})].

A static bias field $\boldsymbol{B}_0= B_0 \hat{\boldsymbol{z}}$, with $\hat{\boldsymbol{z}}$ pointing up, from the source toward the atoms, sets the precession axis. The gyromagnetic ratios of the two manifolds are $\gamma^{(1,2)}= \mp \gamma_0 -\gamma_s$, with $\gamma_0 / 2\pi = 700.98\, \si{kHz/G}$ and $\gamma_s /2\pi = 1.393\, \si{kHz/G}$ \cite{Gomez_2020, Steck_2025}. To first order, and leaving aside nonmagnetic effects that do not depend on the source, such as the Earth's rotation (Sec.~\ref{Subsec: Sequence}), the phases accumulated in a free precession of duration $T$ are
\begin{equation}
\theta^{(f)}=\left(\gamma^{(f)} B_z+c_f \sigma_p\Omega_{p, z}+s_f \Omega_{S, z}\right) T,
\label{Eq: Phases}
\end{equation}
where $B_z$ is the total magnetic field along $\hat{\boldsymbol{z}}$ and $\Omega_S$ is the precession rate that a coupling $g_s^Ng_p^e$ would give a free electron spin [Eq.~(\ref{Eq: Omega_p}) with $m_e$ and $g_p^e$].

In the sum channel, $\theta^{(1)}+\theta^{(2)}$, the electron terms cancel ($s_1+s_2=0$) and the nuclear terms add ($c_1+c_2=2$). Since $\gamma^{(1)}\approx-\gamma^{(2)}$, the response of the sum channel to a magnetic field is reduced to that of the nuclear moment, $2\gamma_s$, a factor $\gamma_0/2\gamma_s\simeq 252$ ($48\, \si{dB}$) below that of either manifold. Weighting each phase instead by the gyromagnetic ratio of the other manifold, as $\gamma^{(2)}\theta^{(1)}-\gamma^{(1)}\theta^{(2)}$, cancels the field term exactly. Divided by $\gamma_0$, this defines our observable, the weighted sum
\begin{equation}
\begin{aligned}
\Theta & \equiv(1-\epsilon)\, \theta^{(1)}+(1+\epsilon)\, \theta^{(2)} \\
& =\left[\left(2-\frac{\epsilon}{2}\right) \sigma_p\Omega_{p, z}+\frac{\epsilon}{2}\, \Omega_{S, z}\right] T,
\end{aligned}
\label{Eq: Weighted sum channel}
\end{equation}
with $\epsilon= \gamma_s /\gamma_0= 1.99 \times 10^{-3}$. Its weights differ from those of the sum channel by only 0.2\%. The weighted sum, however, cancels any magnetic field common to both manifolds, which the sum channel only suppresses. The cancellation is exact to the precision of $\epsilon$, which follows from the $g$ factors of $^{87}$Rb and is known to one part in $10^{6}$ \cite{Steck_2025}.
The price is that $\Theta$ needs the whole turns of each phase (Sec.~\ref{Subsec: Upgrades}). For $\boldsymbol{\srcint}$ along $\hat{\boldsymbol{z}}$, the proton term is $\simeq 2\sigma_p\Omega_p T$, so that a null result bounds $|g_s^Ng_p^p|$.

Through its small electron weight, $\Theta$ also responds to a coupling $g_s^Ng_p^e$ of the source to the electron spin. Since an electron precesses $m_p/m_e$ times faster than a proton for the same coupling, $\Theta$ in fact measures $g_s^N(g_p^p+ 2.7\, g_p^e)$ for $\sigma_p =1/3$. As is customary \cite{Cong_2025, Lee_2018, Kimball_2017}, we quote limits on one coupling at a time. Existing bounds on $g_s^Ng_p^e$ make the electron term negligible at the shortest and longest ranges \cite{Stadnik_2018, Hoedl_2011, Terrano_2015}, but not at ranges of tens of micrometers, where our sensitivity would surpass them \cite{Liang_2023}. There, the sum channel, recorded in the same shots and blind to the electron coupling, can help separate the two couplings once the magnetic field of the source is bounded.

\section{\label{Sec: Sensor} Sensor and source mass}
\subsection{Single-domain spinor condensate}
Our analysis builds on the single-domain $^{87}$Rb spinor Bose-Einstein condensate demonstrated in Refs.~\cite{Gomez_2020, Palacios_2022}. After $4.5\, \si{s}$ of all-optical evaporation in a crossed $1560\, \si{nm}$ optical dipole trap (ODT) of mean frequency $\bar{\omega}= 2\pi \times 90(9)\, \si{Hz}$, about $10^5$ atoms condense in $|F=1, m=0\rangle$, with a condensed fraction above 90\%, in a sphere of Thomas-Fermi radius $\approx 7\, \um$ \cite{Thesis_Palacios_2017, Gomez_2020, Palacios_2022}. All Zeeman sublevels of both manifolds share this spatial mode, so that both manifolds see the same field, free of the gradient shifts that diffusion causes in gas comagnetometers \cite{Sheng_2014}. Within each hyperfine manifold, the atoms form a collective spin $\boldsymbol{F}^{(f)}$. Since the coupling of Eq.~(\ref{Eq: Projection_I}) is linear in the spin, this collective spin precesses at the rate of a single atom, with $\boldsymbol{\srcint}$ averaged over the condensate. An $F=1$ condensate has kept its spin polarization for several seconds, limited by the loss of atoms \cite{Palacios_2018}.

\subsection{Demonstrated performance}\label{Subsec: Demonstrated Performance}
The collective spins can be read out by imaging the magnetization in situ \cite{Vengalattore_2007}, by counting the atoms in each Zeeman sublevel after a Stern-Gerlach separation \cite{Hume_2013}, or by Faraday rotation, which can be quantum nondemolition \cite{SewellNP2013} and repeated during spin precession \cite{Koschorreck_MWM_2010, Palacios_2018}.

We describe the comagnetometer demonstrated in Ref.~\cite{Gomez_2020}, in which both manifolds precess freely and are read out by Faraday rotation. In each shot [Fig.~\ref{Fig: 1c}], an rf $\pi/2$ pulse takes the condensate from $|1,0\rangle$ to an equal superposition of $|1,+1\rangle$ and $|1,-1\rangle$, and a microwave $\pi$ pulse transfers $|1,-1\rangle$ to $|2,-2\rangle$ \cite{Thesis_Gomez_2021}. Each manifold then holds half of the atoms in a stretched state, $F=1$ along the bias field $\boldsymbol{B}_0$ and $F=2$ against it. A second rf $\pi/2$ pulse tips both spins into the plane perpendicular to $\boldsymbol{B}_0$, where they precess freely for a time $T$. Two pulses of an off-resonant probe, propagating perpendicular to $\boldsymbol{B}_0$, then record a few Larmor periods of the transverse spin of each manifold, and fits to these signals give the phases $\theta^{(f)}$ of Eq.~(\ref{Eq: Phases}) \cite{Gomez_2019, Gomez_2020}.

Without a magnetic shield, however, the field changes from shot to shot. In the apparatus of Ref.~\cite{Gomez_2020}, the field along $\boldsymbol{B}_0$ fluctuated by $5.2\, \si{nT}$ rms between shots. The scatter of $\theta^{(1)}$ thus grew as $230(20)\, \si{rad\,s^{-1}}\times T$ and reached $\pi/2$ after only $7\, \si{ms}$, too early to count the whole turns that $\Theta$ needs. The sum channel, whose integer weights make it insensitive to whole turns of either phase, scattered $159(15)$ times less, by $1.45(5)\, \si{rad\,s^{-1}}\times T$, and was the only usable observable.

The readout noise, set by the photon shot noise of the probe, is 898 and 1590 spins for $F=1$ and $F=2$ \cite{Gomez_2019, Thesis_Gomez_2021}. At $T=1\, \si{s}$, with the $2\times10^4$ atoms per manifold that remain of the $5\times10^4$ prepared in each (Sec.~\ref{Sec: sensitivity scenarios}), the readout and projection noise alone would allow a sensitivity of $0.27\, \si{rad\,s^{-1}\,Hz^{-1/2}}$ to the precession rate [Eq.~(\ref{Eq: SO})], for an experimental cycle time $t_{\rm cyc}=20\, \si{s}$. The field that leaks through the sum channel, however, overtakes this noise at $T\approx 17\, \si{ms}$, when all $5\times10^4$ atoms are still present. Since it grows with $T$ as a signal would, the sensitivity saturates at $1.45\, \si{rad\,s^{-1}}\times\sqrt{t_{\rm cyc}}=6.5\, \si{rad\,s^{-1}\,Hz^{-1/2}}$, 24 times worse.

The demonstrated performance highlights three main limitations of the present apparatus: the unscreened laboratory field, a dead time of $19\, \si{s}$ per cycle, spent mostly loading the magneto-optical trap (MOT) from the background vapor \cite{Gomez_2020, Thesis_Palacios_2017, Thesis_Gomez_2021}, and the diffraction of the trap beams, which keeps a source $150\, \um$ below the atoms (Sec.~\ref{Sec: Sources}). Even so, the apparatus as demonstrated would improve on the proton limits between $20\, \um$ and $300\, \um$, by a factor of 50 at $100\, \um$.

\subsection{Upgrades}\label{Subsec: Upgrades}
Three upgrades in preparation would remove these limitations. The first is a magnetic shield, which makes the field predictable enough to use $\Theta$. The readout measures each phase only modulo $2\pi$. For the sum channel this does not matter, since a missed turn of either phase changes $\theta^{(1)}+\theta^{(2)}$ by $2\pi$. The weighted sum, however, can be written as $\Theta=(\theta^{(1)}+\theta^{(2)})-\epsilon\,(\theta^{(1)}-\theta^{(2)})$, and its second term needs the full value of $\theta^{(1)}-\theta^{(2)}$: each miscounted turn shifts $\Theta$ by $2\pi\epsilon$. Because the two manifolds precess in opposite directions, $\theta^{(1)}-\theta^{(2)}$ advances by tens of thousands of turns per second, which the readout cannot count, so their number must be calculated from the field along $\boldsymbol{B}_0$. The field must therefore be known from shot to shot to within $(2\gamma_0 T)^{-1}$, the field that turns $\theta^{(1)}-\theta^{(2)}$ by $1\, \si{rad}$, or $11\, \si{pT}$ at $T=1\, \si{s}$. In finite-element simulations \cite{COMSOL}, a shield of three cylindrical mu-metal layers, with openings for the beams and ports, reduces the drifts of the laboratory field, up to $20\, \si{nT}$ rms \cite{Thesis_Palacios_2017}, to well below this level. Coils inside the shield would set and reverse $\boldsymbol{B}_0$, with a current stable enough to keep it equally predictable.

The second upgrade is a two-dimensional MOT (2D MOT), connected to the science cell through a differential pumping stage. A 2D MOT can deliver up to $9\times 10^9$ atoms per second \cite{Dieckmann_1998}. Even if only a few percent of them were captured, the $2\times 10^8$ atoms of the present MOT \cite{Thesis_Gomez_2021} would load in about $1\, \si{s}$. The differential pumping would also allow a lower background pressure in the science cell.

The third upgrade compresses the trap and probe beams vertically, so that the source can approach the atoms from below [Fig.~\ref{Fig: 1b}, inset]. For example, a $1064\, \si{nm}$ trap and the Faraday probe, both focused at the atoms to vertical waists of $13.5$ and $12\, \um$, would let the apex of the source come within about $33\, \um$ of the atoms. With a margin for alignment tolerances, we take $d_N=50\, \um$ (Sec.~\ref{Sec: Sources}).

\subsection{Source mass}\label{Sec: Sources}
The signal grows with the nucleon density of the source, which favors dense metals such as tungsten, used in many monopole-dipole and spin-velocity searches \cite{Arvanitaki_2014, Aggarwal_2022, Wei_2022}. At $50\, \um$ from the atoms, however, the thermal (Johnson) currents of a metal produce magnetic noise \cite{Varpula_1984, Lee_2008} that the shield cannot screen. Like any static field of the source, including that of a magnetization induced by $\boldsymbol{B}_0$, the slow part of this noise is common to both manifolds and cancels in $\Theta$. Its component at the Larmor frequency, however, tilts the spins, and the quadratic Zeeman shift turns the tilts into noise of $\Theta$. A conductor would also distort the rf pulses and make the state preparation depend on the source position \cite{Wickenbrock_2016}. 

We therefore choose an insulator, bismuth germanate (Bi$_4$Ge$_3$O$_{12}$, BGO), previously used as a source in sub-centimeter searches with noble-gas comagnetometers \cite{Feng_2022, Tullney_2013}. It has a density of $7.13\, \si{g/cm^3}$, is nonmagnetic and transparent at the trap and probe wavelengths, and has no Johnson noise.

Since the trap and probe beams are horizontal, the source sits below the condensate [Fig.~\ref{Fig: 1b}] and must stay clear of them. Each beam widens by diffraction, so that its envelope is a hyperbola. We study a BGO cylinder of radius $R = 9\, \si{mm}$ and height $H = 4\, \si{mm}$ with a spherical top of radius of curvature $R_c = 9.5\, \si{mm}$, $14\, \si{g}$ in all. We require that the source change the intensity of each beam at the condensate by less than $10^{-3}$. Propagating the beams over the source, taken as a mirror, we find that the apex can then come within $134\, \um$ of the atoms with the present $1560\, \si{nm}$ trap beams, and within $35\, \um$ with the refocused beams, in the least favorable case within the alignment tolerances. With a margin of $15\, \um$, we take a gap of $150\, \um$ with the present beams and $d_N=50\, \um$ after the upgrade. A hyperboloid shape, which follows the envelope farther out, would come only $2\, \um$ closer (Appendix~\ref{App: Scans}).

At $50\, \um$, the Casimir-Polder potential of the source, about $h\times0.5\, \si{mHz}$ \cite{Antezza_2004, Obrecht_2007}, is $10^{-8}$ of the trap depth and does not act on the spin. The gap can be calibrated by imaging the condensate together with its mirror image in the polished surface of the source \cite{Harber_2005}.

At the atoms, $\boldsymbol{\srcint}$ is vertical, and since only its component along $\boldsymbol{B}_0$ affects the phases [Eq.~(\ref{Eq: Phases})], $\boldsymbol{B}_0$ is vertical too. For $\lambda\ll R_c$, the source acts as a wall, with $\srcint\to2\pi n\lambda e^{-d_N/\lambda}$, and for $\lambda$ much larger than the source, $\srcint$ no longer depends on $\lambda$. In between, we integrate Eq.~(\ref{Eq: Pseudomagnetic Field}) over the source.

\subsection{Measurement sequence}\label{Subsec: Sequence}
Each shot prepares the two spins, lets them precess for a time $T$, and reads out both phases, which give $\Theta$ [Fig.~\ref{Fig: 1c}]. Between shots, the source alternates between two positions: in, with its apex $d_N$ below the atoms, and out, $40\, \si{mm}$ toward the side port, where $\srcint$ is less than 1\% of its in-position value. Moving it adds $0.5\, \si{s}$ to the mean cycle. The sign of $\boldsymbol{B}_0$ is reversed between blocks of in-out pairs, and the in-out difference of $\Theta$, averaged over both signs,
\begin{equation}
\Delta\Theta=\tfrac12\big[(\Theta_{+,\rm in}-\Theta_{+,\rm out})+(\Theta_{-,\rm in}-\Theta_{-,\rm out})\big],
\label{Eq: Estimator}
\end{equation}
keeps only the terms that, like the signal, change with the source position and do not reverse with $\boldsymbol{B}_0$.

This estimator removes three classes of systematic effects by symmetry. The weights of $\Theta$ cancel the Larmor precession in any magnetic field common to both manifolds: drifts of the bias field, the laboratory field that leaks through the shield, and any field that moves with the source. The in-out difference cancels every phase that does not depend on the source position, such as those from the Earth's rotation, light shifts, the quadratic Zeeman shift, collisions, and pulse or readout errors, along with their slow drifts. The average over both signs of $\boldsymbol{B}_0$ cancels every phase that reverses with $\boldsymbol{B}_0$, which includes all phases of the spin dynamics. A systematic difference between in and out shots in the atom number, the density, or a pulse amplitude therefore cancels as well. Only effects that, like the signal, depend on the source position and do not reverse with $\boldsymbol{B}_0$ survive, such as a rotation of the rf polarization axis or of the probe direction about $\hat{\boldsymbol{z}}$, a change in the circular polarization of the rf and microwave fields, and light that the source reflects toward the atoms. Imperfect weights or reversal leave residuals of what they cancel, and errors in the gap or in the density rescale the limit but cannot mimic the signal. A full assessment of systematic effects will require the experiment.

\section{\label{Sec: Sensitivity Reach} Projected sensitivity}
\subsection{Sensitivity scenarios}\label{Sec: sensitivity scenarios}
Each cycle, of duration $t_{\rm cyc}$, measures $\Theta/T$ with a noise $\delta \Theta/T$, so that after a time $t$ the resolution is $S_{\Omega}/\sqrt{t}$, with
\begin{equation}
\begin{aligned}
S_{\Omega} & =\frac{\delta \Theta \sqrt{t_{\rm cyc}}}{T}, \\
\left(\delta \Theta \right)^2 & =\sum_{f=1,2}w_f^2\left[\left(\frac{\delta F_{\rm RO}^{(f)}}{f N}\right)^2+\frac{1}{2 f N}\right],
\end{aligned}
\label{Eq: SO}
\end{equation}
where $w_{1,2}= 1 \mp \epsilon$ are the weights of $\Theta$ and $N$ is the number of atoms per manifold at the readout. Each manifold carries a collective spin of length $\hbar fN$, whose azimuth is read out with an uncertainty $\delta F_{\rm RO}^{(f)}/fN$ from the readout noise and $1/\sqrt{2fN}$ from the projection noise of a coherent spin state \cite{Gomez_2020}. The magnetic field does not enter, since $\Theta$ cancels it and the shield keeps it predictable enough to count the precession cycles (Sec.~\ref{Subsec: Upgrades}). Table~\ref{Tab: Sensitivitty scenarios} lists the parameters of the two scenarios.

\textit{Scenario U (upgraded)} adds the three upgrades of Sec.~\ref{Subsec: Upgrades}, so that $d_N= 50\, \um$. It assumes $N=2\times10^4$ atoms per manifold at the readout, what remains of the initial $5\times10^4$ after $1\, \si{s}$, and the demonstrated readout noise \cite{Gomez_2019, Thesis_Gomez_2021}, although the refocused probe, with a smaller cross section, would couple more strongly to the atoms. Lowering the bias field from the $120\, \si{mG}$ of Ref.~\cite{Gomez_2020} to $60\, \si{mG}$ reduces the quadratic Zeeman shift fourfold. Our truncated-Wigner simulations \cite{Blakie_2008, Polkovnikov_2010} of both manifolds, which include the spin-dependent collisions and the losses, then give $12\, \si{mrad}$ of twisting noise at $T=1\, \si{s}$. We add it in quadrature to the readout noise, which is five times larger. The cycle lasts $t_{\rm cyc}=8.0\, \si{s}$. Besides the precession time $T=1\, \si{s}$, it includes $1\, \si{s}$ of loading from the 2D MOT, the $5.35\, \si{s}$ of ODT loading, hold and evaporation of the present sequence \cite{Thesis_Palacios_2017, Gomez_2020}, $0.15\, \si{s}$ for state preparation and readout, and $0.5\, \si{s}$ for moving the source.

\textit{Scenario Q (quantum-limited)} assumes the projection-noise limit, which needs three further changes. The first is a longer lifetime. Collisions between $F=1$ and $F=2$ atoms limit the lifetime of the mixture to about $1\, \si{s}$ at the present density. A softer trap, at $\bar{\omega}= 2\pi \times 30\, \si{Hz}$, would lower the density and extend this lifetime to about $8\, \si{s}$ \cite[Sec.~6.5]{Thesis_Gomez_2021}, and the lower background pressure of the upgraded vacuum would lengthen the one-body lifetime, at present 8--$12\, \si{s}$ \cite{Palacios_2018, Palacios_2022}. As a best case, we assume that the better vacuum makes one-body losses negligible, so that the lifetime for all losses together is $8\, \si{s}$. With $T=8\, \si{s}$, $1.8\times10^4$ of the initial $5\times10^4$ atoms per manifold then remain at the readout. The second change is a readout noise well below the projection noise, as demonstrated with cavity-enhanced dispersive probing \cite{Hosten_2016}, interaction-based readout \cite{Linnemann_2016, Hosten_2016_Science}, and fluorescence counting with single-atom resolution \cite{Hume_2013}. The third is the suppression of the twisting, which over $8\, \si{s}$ would exceed the projection noise several times. The twisting turns the random longitudinal spin of each manifold, set by the projection noise, into a random phase. It has two causes: the quadratic Zeeman shift, which microwave dressing can cancel \cite{Gomez_2020}, and the collisions, which shift the precession frequency of each manifold in proportion to the longitudinal spins of both manifolds. These spins stay constant during the free precession, apart from losses. Measuring them beforehand, nondestructively and below the projection noise \cite{Koschorreck_MWM_2010, Sewell_2010, Hosten_2016}, gives the angle that the collisions add in each shot, which can then be subtracted.

\begin{table}[t]
\caption{\label{Tab: Sensitivitty scenarios}Sensitivity scenarios U (upgraded) and Q (quantum-limited). Upper rows: inputs, with the atom number $N$ counted at the readout. Lower rows: per-shot noise of $\Theta$, sensitivity [Eq.~(\ref{Eq: SO})], and limit at the 95\% confidence level (C.L.) after one week.}
\begin{ruledtabular}
\footnotesize\setlength{\tabcolsep}{2.5pt}
\begin{tabular}{lcc}
 & U & Q\\
\hline
Bias field $B_0$ (mG) & 60 & 60\\
Trap frequency $\bar\omega/2\pi$ (Hz) & 90 & 30\\
Gap $d_N$ ($\um$) & 50 & 50\\
Precession time $T$ (s) & 1 & 8\\
Cycle time $t_{\rm cyc}$ (s) & 8.0 & 15.0\\
Atoms per manifold $N$ & $2\times10^4$ & $1.8\times10^4$\\
Readout noise $\delta F_{\rm RO}^{(1,2)}$ (spins) & 898, 1590 & $\ll\sqrt{fN/2}$\\[2pt]
\hline\noalign{\vskip 2pt}
Per-shot noise $\delta\Theta$ (mrad) & 61 & 6.4\\
Sensitivity $\SO$ (rad\,s$^{-1}$\,Hz$^{-1/2}$) & 0.17 & 0.0031\\
$|g_s^Ng_p^p|_{\min}$ at $\lambda=100\,\um$ & $2{\times}10^{-22}$ & $3{\times}10^{-24}$\\
\end{tabular}
\end{ruledtabular}
\end{table}

\subsection{Projected reach}\label{Subsec: reach}
\begin{figure}[t]
\centering
\includegraphics[width=\columnwidth]{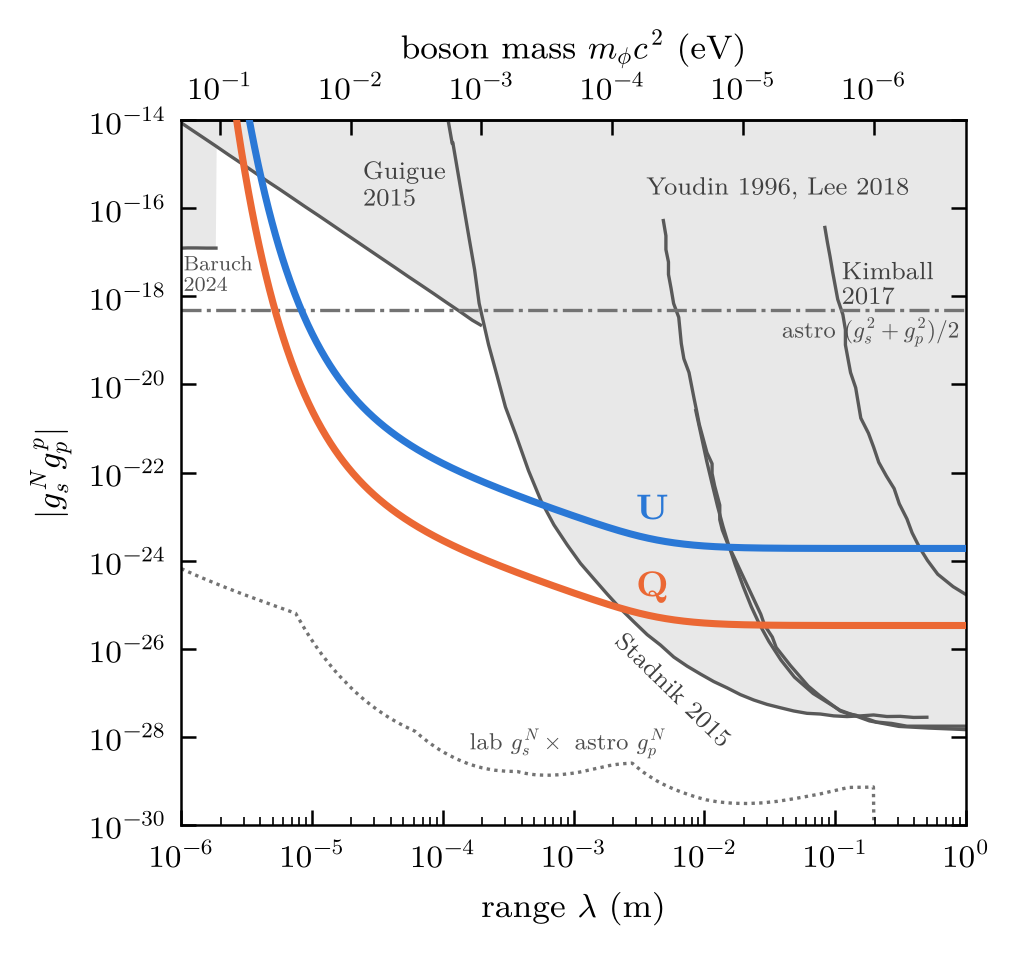}
\caption{\label{Fig: constraints}Projected 95\% C.L. sensitivity to the proton coupling after one week, for scenarios U (blue) and Q (orange). Gray: existing laboratory limits at their published confidence levels, labeled by reference: Guigue 2015 and Lee 2018 ($^3$He) \cite{Guigue_2015, Lee_2018}, Stadnik 2015 ($^{129}$Xe and $^3$He data of Ref.~\cite{Tullney_2013}) \cite{Stadnik_2015}, Youdin 1996 (Cs) \cite{Youdin_1996, Kimball_2015}, Kimball 2017 (Rb) \cite{Kimball_2017}, and Baruch 2024 (molecular electric dipole moments) \cite{Baruch_2024}. Dash-dotted and dotted lines: indirect bounds (Sec.~\ref{Subsec: reach}) \cite{Bottaro_2023, Carenza_2019, Raffelt_2012}. Curves from the data release \cite{Cong_data} of Ref.~\cite{Cong_2025}, except Rb and Cs, which are taken from Ref.~\cite{Kimball_2017} and rescaled to our convention.}
\label{Fig: Constraints}
\end{figure}

With acquisition cycles split equally between the two source positions, the in-out difference of Eq.~(\ref{Eq: Estimator}) measures the precession rate with an uncertainty $\delta\Omega=2S_{\Omega}/\sqrt{t}$. The proton term of $\Theta$ is $\simeq 2\sigma_p\Omega_p T$ (Sec.~\ref{Subsec: Observable}), with $\Omega_p$ from Eq.~(\ref{Eq: Omega_p}). At the 95\% confidence level (C.L., $1.96\sigma$), as in most of the searches we compare with, a null result bounds the coupling by
\begin{equation}
|g_s^Ng_p^p|_{\min}(\lambda)=\frac{4 \pi m_p}{\hbar}\, \frac{1.96\,\delta\Omega}{2 \sigma_p \srcint(\lambda)}.
\label{Eq: Kmin}
\end{equation}
Figure~\ref{Fig: Constraints} shows the result for one week of acquisition, and Appendix~\ref{App: Scans} shows how other conventions change it. At $\lambda=100\, \um$, $|g_s^Ng_p^p|_{\min}\approx 2\times10^{-22}$ and $3\times10^{-24}$ in scenarios U and Q, $5\times10^3$ and $3\times10^5$ times below the best existing laboratory limit, inferred from $^{3}$He \cite{Guigue_2015, Cong_2025} at 95\% C.L.

The two scenarios improve on the existing proton limits for $\lambda$ between $4\, \um$ and $0.6\, \si{mm}$ (boson masses of 0.3--$50\, \si{meV}/c^2$) and between $3\, \um$ and $2\, \si{mm}$ (0.09--$70\, \si{meV}/c^2$), with the largest gains, $8\times10^3$ and $4\times10^5$, at $\lambda=0.2\, \si{mm}$, where the compiled $^3$He limit \cite{Guigue_2015, Cong_2025} ends. At longer range, the limit reinterpreted \cite{Stadnik_2015, Cong_2025} from the $^{129}$Xe and $^3$He data of Ref.~\cite{Tullney_2013} is stronger, but it rests on proton spin contents that differ between nuclear models by more than an order of magnitude \cite{Kimball_2015, Cong_2025}. In $^{87}$Rb, by contrast, the Schmidt value of $\sigma_p$ is reliable to within a factor of two, and the neutrons carry little of the nuclear spin (Sec.~\ref{Subsec: Signal}).

Figure~\ref{Fig: Constraints} also shows indirect bounds, which combine separate limits on the two couplings. Stars would lose energy by emitting light bosons through either coupling, so that white dwarfs bound the scalar coupling \cite{Bottaro_2023} and the neutrino burst of SN\,1987A bounds the pseudoscalar one \cite{Carenza_2019}. The compilation of Ref.~\cite{Cong_2025} combines the two as $(g_s^2+g_p^2)/2$ (dash-dotted). The scalar coupling, however, is bounded far more tightly in the laboratory, by short-range tests of gravity, because the same boson would add a Yukawa force, proportional to $(g_s^N)^2$, between two unpolarized masses. The strongest indirect bound therefore pairs this laboratory bound on the scalar coupling with the supernova bound on the pseudoscalar one \cite{Raffelt_2012}. It is the dotted curve, far below both scenarios.

These indirect bounds assume that the boson couples in the same way in a supernova core, in a stellar plasma and in the laboratory. Couplings that depend on the environment evade them \cite{Jaeckel_2007, DeRocco_2020}, and the supernova bound also depends on how SN\,1987A is modeled \cite{Bar_2020}. A laboratory search makes none of these assumptions.

\section{\label{Sec: Conclusion} Discussion and outlook}
Below $5\, \si{mm}$, the monopole-dipole coupling of the proton has so far been bounded mainly through nuclear models of neutron-dominated nuclei. A $^{87}$Rb condensate, whose nuclear spin is carried by a valence proton, offers a direct test. Trapped by light in vacuum, it can sit $50\, \um$ from a dense insulating source without an intervening cell wall, and a weighted sum of the phases of its two hyperfine manifolds cancels the magnetic field common to both. With the three upgrades of Sec.~\ref{Subsec: Upgrades}, one week of data would improve on the existing limits from $4\, \um$ to $0.6\, \si{mm}$, by up to $8\times10^3$. The main limit would then be the spin readout, at present ten times above the projection noise. At the projection-noise limit, the improvement would extend from $3\, \um$ to $2\, \si{mm}$, by up to $4\times10^5$.

Beyond this search, the same apparatus could address other questions, each of which would need a dedicated study. Without a nearby source in vacuum, the Earth acts as a large but distant one \cite{Venema_1992, Kimball_2017, Zhang_2023, Agrawal_2023}. A long-range monopole-dipole interaction with its nucleons, or a coupling $\boldsymbol{\sigma}\cdot\boldsymbol{g}$ of the proton spin to gravity, which general relativity forbids, would shift $\Theta$ by a constant that, unlike the offsets of the spin dynamics, survives the reversal of $\boldsymbol{B}_0$. With no in-out difference to cancel them, however, all other offsets fixed in the laboratory would then have to be controlled in absolute terms.

Sources with different neutron-to-proton ratios would test whether $g_s^p=g_s^n$. The same data would also bound the electron coupling $g_s^Ng_p^e$ without any nuclear model, since the electron weight of $\Theta$ is fixed by the magnetic moments alone. They would improve on the existing laboratory limits at ranges of tens of micrometers.

A moving source would probe the velocity-dependent potentials $V_{4+5}\propto \boldsymbol{\sigma}\cdot (\boldsymbol{v}\times \hat{\boldsymbol{r}})$ and $V_{12+13}\propto\boldsymbol{\sigma}\cdot \boldsymbol{v}$ \cite{Dobrescu_2006, Fadeev_2019, Cong_2025}, which for the proton have been tested with macroscopic sources only above a few centimeters \cite{Wei_2022, Shu_2024}. The source would have to move fast, in vacuum and close to the atoms, throughout the precession. A source spinning below the condensate, for example, would keep a fixed gap and, if anisotropic, would also modulate the static signal at a frequency of choice. Reversing its rotation would reverse the velocity-dependent signals but not the static one.

Building on the demonstrated condensate comagnetometer \cite{Gomez_2020}, we have analyzed the prospects for a direct search for the monopole-dipole coupling of the proton at submillimeter ranges, shown that it is feasible, and projected a sensitivity orders of magnitude beyond the existing limits. At this sensitivity, the search could even reveal a new interaction. A detection would point to a boson far lighter than any known elementary scalar and to a new source of $\CP$ violation, with profound consequences for physics beyond the Standard Model and for the nature of dark matter.

\appendix
\section{Dependence on the design parameters}\label{App: Scans}

\begin{figure*}[tp]
\centering
\includegraphics[width=0.9\textwidth]{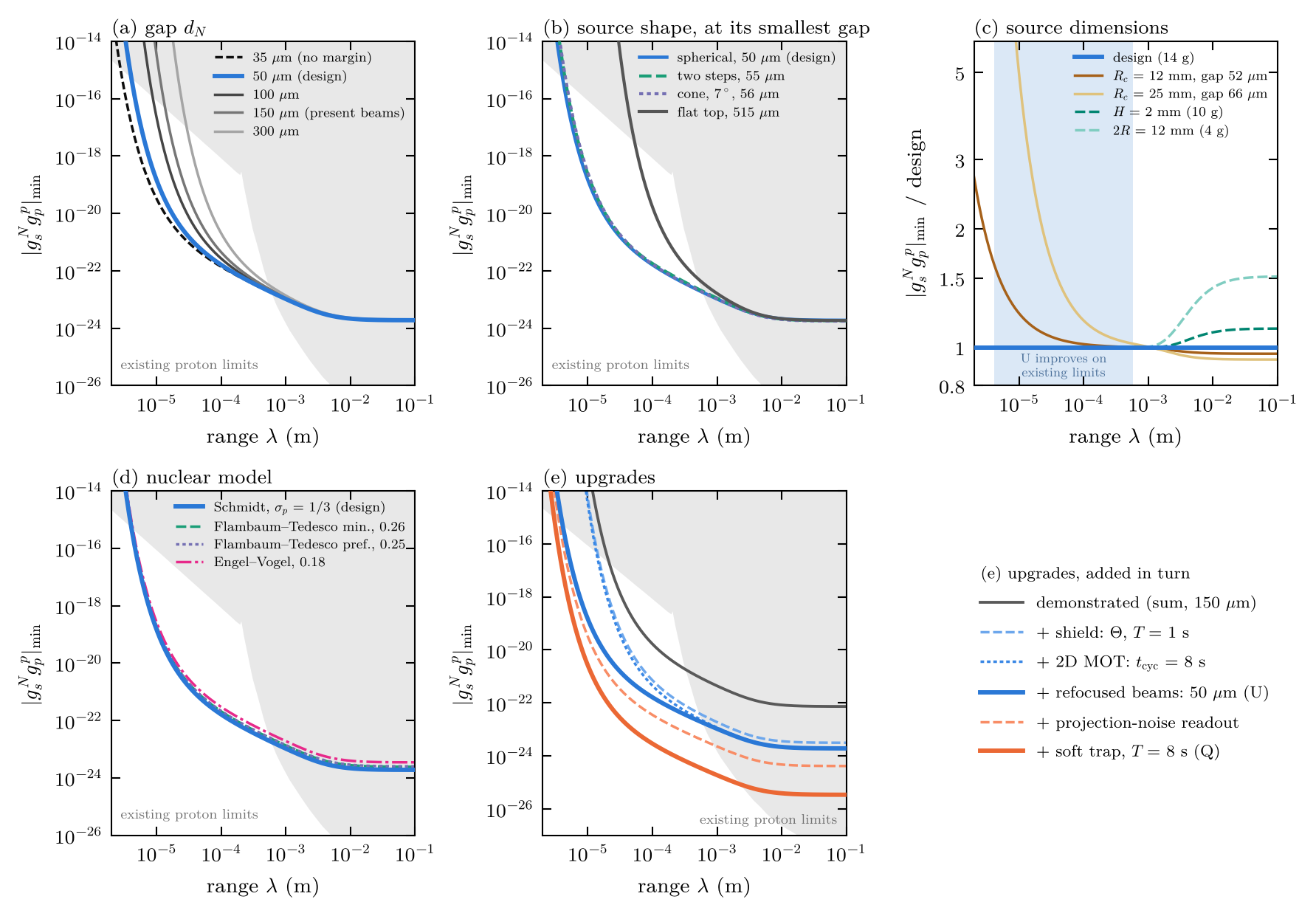}
\caption{\label{Fig: Scans}Projected reach of scenario U (one week, 95\% C.L.) as one design choice is varied at a time, with the existing proton limits in gray. (a)~Gap $d_N$: $35\, \um$ is the clearance of the refocused beams, without margin, and $150\, \um$ the gap with the present beams. (b)~Shape of the top [Fig.~\ref{Fig: Sources}(a)], each placed $15\, \um$ beyond its own clearance. (c)~Dimensions of the source [Fig.~\ref{Fig: Sources}(b)], relative to the design, with the window of U shaded. (d)~$\sigma_p$ of the Schmidt model (design) and of semi-empirical models \cite{Flambaum_2006, Engel_1989}. (e)~Upgrades added in turn, from the apparatus as demonstrated to scenario Q.}
\end{figure*}

\begin{figure*}[tp]
\centering
\includegraphics[width=\textwidth]{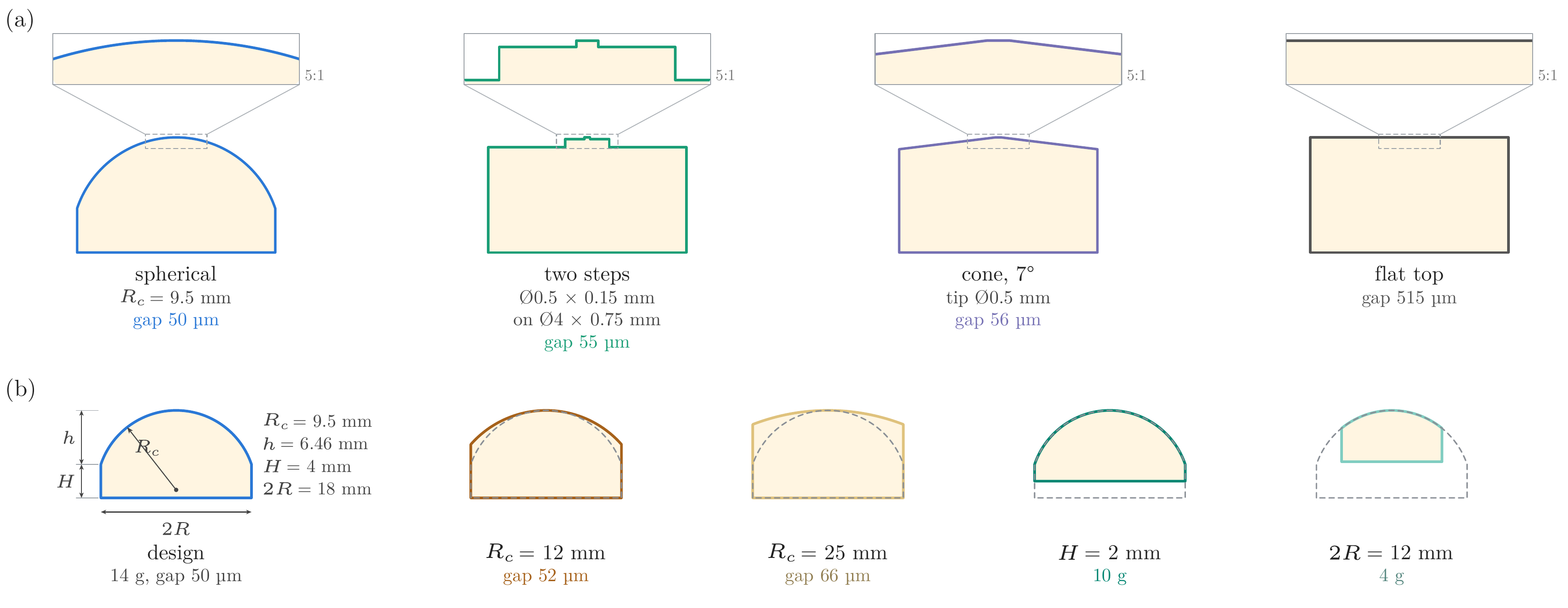}
\caption{\label{Fig: Sources}Sources of Fig.~\ref{Fig: Scans}, in cross section and to scale. (a)~The four shapes of Fig.~\ref{Fig: Scans}(b), all with the base and height of the design, each with its gap and a 5:1 detail of its apex. (b)~The design, $14\, \si{g}$ of BGO, and the variants of Fig.~\ref{Fig: Scans}(c), each drawn over the design (dashed gray).}
\end{figure*}

Figures~\ref{Fig: Scans} and \ref{Fig: Sources} show how the reach of scenario U changes when one design choice is varied at a time. By Eq.~(\ref{Eq: Kmin}), the limit is proportional to the uncertainty $\delta\Omega$ of the precession rate and inversely proportional to $\sigma_p\srcint$. The geometry of the source enters only through $\srcint$ and sets how the limit depends on $\lambda$. The sensor and the nuclear model rescale it at every range.

\textit{Gap.} The signal comes mostly from the part of the source within about $\lambda$ of the atoms. Once $\lambda$ is shorter than the gap, the signal therefore falls as $e^{-d_N/\lambda}$, and the gap sets the shortest range we can reach [Fig.~\ref{Fig: Scans}(a)]. Because the reach at longer range beats the existing limits by orders of magnitude, the window still extends down to about a tenth of the gap: $4\, \um$ for $d_N=50\, \um$, and $14\, \um$ with the present beams ($d_N=150\, \um$).

\textit{Shape and size.} The beams pass just above the apex and widen away from it, so the best surface falls away from them at least as fast as they widen. A spherical top does so near the apex. Of the four shapes of Fig.~\ref{Fig: Scans}(b), each placed $15\, \um$ beyond its own clearance, it comes closest. A two-step top or a cone sits about $5\, \um$ farther away, which matters only at the shortest ranges. A flat top would cut the beams unless it sat $0.5\, \si{mm}$ away, and it would improve on the existing limits only above about $50\, \um$. A hyperboloid, which follows the beams farther out, could come $2\, \um$ closer than the spherical top, a small gain. A flatter spherical top must sit farther away [Fig.~\ref{Fig: Scans}(c)]. Material more than about $\lambda$ from the atoms contributes little, so the overall size of the source matters only above $\lambda\approx1\, \si{mm}$, outside the window of U.

\textit{Inputs that rescale the limit.} Other inputs scale the limit by the same factor at every range. The semi-empirical values of $\sigma_p$ would weaken it 1.3--1.8 times [Fig.~\ref{Fig: Scans}(d)], and a 90\% C.L. would strengthen it 1.2 times. Scenario U is limited by the readout, so halving the readout noise or doubling the atom number would improve it nearly twofold. In scenario Q, the lifetime matters most, because it limits both the precession time and the atoms left at the readout: $4\, \si{s}$ instead of $8\, \si{s}$ would weaken the limit 1.7 times.

\textit{Upgrades.} Each upgrade removes one limitation [Fig.~\ref{Fig: Scans}(e)]. At $\lambda=100\, \um$, the shield gains the most, a factor of 23, because it makes $\Theta$ usable and removes the field noise that capped the sum channel. The 2D MOT shortens the cycle and gains 1.6. The refocused beams bring the source closer and gain 2.7, and 28 at $30\, \um$, where the gap matters most. A readout below the projection noise gains 4.6. The softer trap, with $T=8\, \si{s}$ and the twisting suppressed, gives the last factor of 12, to scenario Q.

\begin{acknowledgments}
This work was funded by the European Commission projects Field-SEER (ERC 101097313), OPMMEG (101099379) and QUANTIFY (101135931), by the Spanish Ministry of Science (MCIN) projects SAPONARIA (PID2021-123813NB-I00) and SALVIA (PID2024-158479NB-I00) and the ``Severo Ochoa'' Center of Excellence CEX2019-000910-S, by the Generalitat de Catalunya through the CERCA program, DURSI grant No.~2021~SGR~01453 and QSENSE (GOV/51/2022), and by Fundaci\'{o} Privada Cellex and Fundaci\'{o} Mir-Puig.

IFAE is partially funded by the CERCA program of the Generalitat de Catalunya.
This publication is part of the R\&D\&i project PID2023-146686NB-C31 funded by MICIU/AEI/10.13039/501100011033/ and by ERDF/EU.
D.~Blas acknowledges financial support from the Spanish Ministry of Science and Innovation (MICINN) through the Spanish State Research Agency, under Severo Ochoa Centres of Excellence Programme 2025-2029 (CEX2024-001442-S).
This work is supported by ERC grant (GravNet, ERC-2024-SyG 101167211, DOI: 10.3030/101167211). Funded by the European Union. Views and opinions expressed are however those of the author(s) only and do not necessarily reflect those of the European Union or the European Research Council Executive Agency. Neither the European Union nor the granting authority can be held responsible for them.
D.~Blas acknowledges the support from the European Research Area (ERA) via the UNDARK project of the Widening participation and spreading excellence programme (project number 101159929).
\end{acknowledgments}

\section*{Data availability}
The existing limits shown in Fig.~\ref{Fig: Constraints} are openly available \cite{Cong_data}. The code and data behind the projections are available from the authors upon reasonable request.

\clearpage
\bibliography{bib_new_new}

\end{document}